\documentclass[]{spie}  

\usepackage{amsmath,amsfonts,amssymb}
\usepackage{graphicx}
\usepackage[colorlinks=true, allcolors=blue]{hyperref}
\usepackage[perpage]{footmisc} 
\usepackage{nicefrac}
\usepackage{subcaption}
\usepackage{csquotes}
\usepackage{xspace}

\def\corgidb{\texttt{corgidb}\xspace}
\def\corgietc{\texttt{corgietc}\xspace}

\newcommand{\reffig}[1]{Figure \ref{#1}}
\newcommand{\refsec}[1]{Section \ref{#1}}
\newcommand{\refnum}[1]{Ref.~\citenum{#1}}
\newcommand{\nlhref}[1]{\href{#1}{\nolinkurl{#1}}} 

\title{The Roman Coronagraph Community Participation Program: target database and tools}

\author[a,b,c]{Dmitry Savransky}
\affil[a]{Sibley School of Mechanical and Aerospace Engineering, Cornell University, Ithaca, NY, 14853, USA}
\affil[b]{Carl Sagan Institute, Cornell University, Ithaca, NY, 14853, USA}
\affil[c]{Department of Astronomy, Cornell University, Ithaca, NY, 14853, USA}

\author[a]{Rifah Tasnim}
\author[c]{Saanika Choudhary}
\author[a]{Hovik Grigoryan}
\author[d]{Justin Hom}
\author[e]{Bijan Nemati}
\author[f]{Neil Zimmerman}
\author[d]{Ramya Anche}
\author[a]{Nicholas Phillips}
\author[a]{Louie Donesa}
\author[a]{Parth Mittal}
\author[a]{Amira Razack}
\author[a]{Savaas Iqbal}
\author[a]{Javier Majumdar}
\author[d]{Schuyler G. Wolff}
\author[g]{Vanessa P. Bailey}

\affil[d]{Steward Observatory and the Department of Astronomy, The University of Arizona, 933 N Cherry Ave, Tucson, AZ, 85721, USA}
\affil[e]{Tellus1 Scientific, LLC, Madison, AL, 35756, USA}
\affil[f]{NASA Goddard Space Flight Center, 8800 Greenbelt Rd., Greenbelt, MD 20771, USA}
\affil[g]{Jet Propulsion Laboratory, California Institute of Technology, 4800 Oak Grove Drive, Pasadena, CA 91109, USA}

\authorinfo{Send correspondence to Dmitry Savransky 
\texttt{ds264@cornell.edu}}

\begin{document} 
\maketitle

\begin{abstract}
The Nancy Grace Roman Space Telescope, set to launch in Fall 2026, will carry the Coronagraph Instrument, which will, for the first time, demonstrate high-contrast imaging with active wavefront control in visible wavelengths from space. In preparation for execution of the Coronagraph’s commissioning and observing programs, the Roman Coronagraph Community Participation Program (CPP) has developed a target database and associated ecosystem of publicly accessible tools for observation planning and scheduling. The target database includes both stars and known sub-stellar companions and disks that may be observed by the Coronagraph instrument during its primary mission. Targets in the database include planet and disk hosts as well as calibration stars, reference stars, and engineering program targets. The database is designed to operate in conjunction with a variety of tools, including an exposure time calculator, a pointing and keepout calculator, and a reference star selection tool.  Here, we describe the current schema and contents of the database and demonstrate how it and its associated tools are being used for observation planning.
\end{abstract}


\section{INTRODUCTION}\label{sec:intro} 
The Nancy Grace Roman Space Telescope (Roman) is NASA's next flagship astrophysics observatory, with a current launch readiness date of August 30th, 2026. Roman has a 2.4 m near-infrared-optimized telescope and will carry the Wide-Field Instrument and
a Coronagraph Instrument technology demonstration.  The Coronagraph Instrument will, for the first time, demonstrate in-space use of several key technologies required for the direct imaging of exoplanets.  These include ultra-precise wavefront sensing and control, large-format deformable mirrors, high-contrast coronagraphs designed for obscured and complex primary mirrors, and photon-counting, EMCCD detectors. The Coronagraph Instrument data will also allow for testing of advanced data post-processing techniques in a contrast regime where they have never been previously applied.

Extensive, detailed descriptions of the Roman Observatory and the Coronagraph Instrument are available throughout the literature, and especially in Refs.~\citenum{spergel2015wide,akeson2019wide}. An extensive review of past work on simulation and performance modeling for the Coronagraph Instrument can be found in \refnum{douglas2020review}.  A definitive description of the coronagraph design is available in \refnum{kuan2025roman} and its optical modeling is described in \refnum{krist2023end}.  A complete description of the coronagraph flight masks can be found in \refnum{riggs2025flight}, and further details on the instrument's spectroscopic and polarimetric capabilities are available in \refnum{groff2025spectroscopy}. The analytical performance model and error budget for the Coronagraph Instrument have been published in \refnum{nemati2023analytical}.  The results of the instrument-level test program are available in \refnum{poberezhskiy2025overview}, and the high-order wavefront sensing and control architecture are presented in Refs.~\citenum{cady2025high, zhou2025high}.

The Coronagraph Instrument will be capable of imaging giant planets in reflected light in visible wavelengths, and will potentially be sensitive to true Jovian analogues.  The Coronagraph observing program will include exoplanets previously discovered via doppler spectroscopy as well as young, giant planets previously imaged in thermal emission with ground-based instruments. For details on the observing program see Refs.~\citenum{wolff2024roman,wolff2026roman}.  The Roman coronagraph's default mode of operations for coronagraphic observations will include integrating on two separate stars: the target star and a reference star.  The reference star is used for high-order wavefront sensing and control (to iteratively create the region of high contrast via focal plane wavefront sensing) and subsequently in post-processing of the target star data.  The reference star must therefore be bright (V magnitude $<3$), unresolved (angular diameter of $< 2$ mas), and have no companions or disks of its own, which limits possible references to fewer than 100 stars.  For more details on reference star selection, see \refnum{hom2026roman}.

The Roman Coronagraph Community Participation Program (CPP) was established by the 2022 NASA ROSES Nancy Grace Roman Space Telescope Research and Support Participation Opportunities program element. The CPP category of this call was limited to sole-PI or small-group proposals, with the intent to form a single, cohesive CPP Team from the selected teams, as well as international partners and representatives from the Coronagraph Instrument science and engineering teams from NASA's Jet Propulsion Laboratory and Goddard Space Flight Center, and members of the Roman Science Support Center (SSC) at Caltech/IPAC (an element of the Roman Ground System). Seven US-based teams were selected via the initial ROSES program element, with 3 additional teams added via a subsequent solicitation. A complete overview of the CPP can be found in \refnum{savransky2024roman}. Since 2023 the CPP has been developing tools for processing and simulating coronagraph data (described in detail in \refnum{wang2026roman}) and in support of planning the instrument's calibration and observing programs.  This latter set of tools is designed around a database of coronagraph targets and auxiliary target data. Here, we describe the schema and design of this database, and the toolset now being used for observation planning. 

\section{corgidb: The Roman Coronagraph Target Database}\label{sec:corgidb}

\begin{figure}[ht]
\centering
\includegraphics[width=0.9\textwidth]{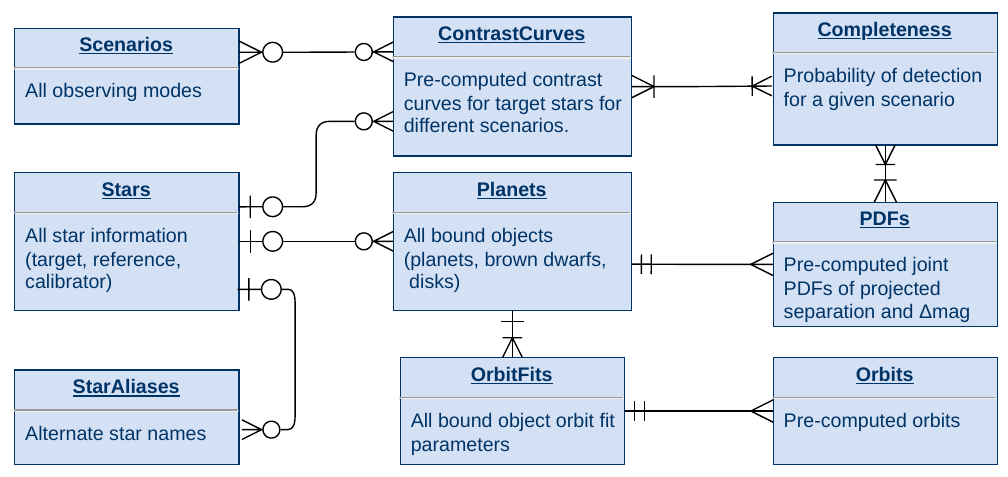}
\caption{\corgidb schema. The database is composed of multiple tables, shown here schematically with their relationships.}\label{fig:corgidb_schema}
\end{figure}

\corgidb\footnote{\nlhref{https://github.com/roman-corgi/corgidb}} provides target and auxiliary information for observation planning, and interfaces directly with various planning tools. The database design is derived from the Imaging Mission Database (\nlhref{https://plandb.sioslab.com}), which was a product from the WFIRST Science Investigation Teams and tracked potential coronagraph targets based on the CDR-era instrument design.\cite{savransky2019exploration} \corgidb  is implemented in MySQL and is composed of multiple tables, shown schematically in \reffig{fig:corgidb_schema}.  The primary table is the \texttt{Stars} table, which contains unique entries for every star that will potentially be observed with the Roman coronagraph.  This includes not only target stars relevant to the coronagraph's observing program, but also all reference stars, calibration stars, and engineering targets.  An auxiliary \texttt{StarAliases} table provides alias data for all objects in the Stars table, allowing for automated name resolution.  The aliases are composed of all entries available in the SIMBAD Astronomical Database, the NASA Exoplanet Archive, and internal naming conventions used by the CPP.

All known companion objects of interest are stored in the \texttt{Planets} table.  This includes planetary-mass objects as well as brown dwarfs, other stellar companions, and disks. Each `planet' object can be associated with one or more orbital fits (stored in table \texttt{OrbitFits}).  One fit is defined as the default (i.e., used for planning purposes) and the companion object's position with respect to its host star is evaluated over the course of one full orbital period, with the results stored in the \texttt{Orbits} table. For each of the coronagraph's observing modes, we define an operating scenario, composed of the mode, the target signal-to-noise ratio (SNR) for a successful observation, and limits on allowable integration times on target. These are encoded in the \texttt{Scenarios} table.  For each star and scenario, a contrast curve is generated using \corgietc (see \refsec{sec:corgietc}) and stored in the \texttt{ContrastCurves} table.  Given the contrast curve and joint density function of planet brightness and projected separation (see Refs.~\citenum{brown2005,savransky2019exploration} for details), we can compute the probability of detection, which is stored in the \texttt{Completeness} table. 

Access to the database is provided via either a generic SQL query interface, or via pre-composed queries implemented in PHP wrappers which return standard, JSON-formatted database outputs.  The \corgidb backend provides Python-based wrappers for these pre-composed queries, allowing them to be integrated with any tool written in Python.  Currently, the queries include a star name resolver, returning information on a single entry in the \texttt{Stars} table, and returning all entries in the \texttt{Stars} table that are labeled as reference stars.

\section{corgietc: The Roman Coronagraph Exposure Time Calculator}\label{sec:corgietc}
\corgietc\footnote{\nlhref{https://github.com/roman-corgi/corgietc}} is the coronagraphic exposure time calculator for point source detections with post-processing.  This tool is designed to answer three questions:
\begin{enumerate}
\item For a given observing mode and assumptions on the brightness of astrophysical noise sources, what is the minimum achievable flux ratio ($F_r$ - the ratio of planet to star flux) detectable at a given SNR for a given integration time?
\item For a given observing mode, what is the minimum achievable flux ratio detectable given infinite integration time (i.e., what is the effective noise floor of the system for this mode)?
\item For a given observing mode and assumptions on the brightness of astrophysical target and noise sources, what is the integration time required to reach a given SNR in post-processed coronagraphic images?
\end{enumerate}
\corgietc is built on top of the EXOSIMS framework\cite{savransky2017exosims} as an EXOSIMS \texttt{OpticalSystem} implementation\footnote{\nlhref{https://exosims.readthedocs.io/en/latest/opticalsystem.html}}. The code implements the instrument model described in \refnum{nemati2023analytical}.

\corgietc currently implements optimistic and conservative versions of four observing modes:
\begin{enumerate}
\item Imaging (IMG) in a narrow field of view (NF) in Band 1 (centered at $\sim575$ nm) with a Hybrid Lyot Coronagraph (HLC)---inner working angle (IWA) 3 $\lambda/D$, outer working angle (OWA) 9.7 $\lambda/D$
\item Imaging in a wide field of view (WF) in Band 4 (centered at $\sim825$ nm) with a Shaped Pupil Coronagraph (SPC)---IWA 5.9 $\lambda/D$, OWA 20.1 $\lambda/D$
\item Spectroscopy (SPEC) in a narrow field of view in Band 3 (centered at $\sim730$ nm) with an SPC---IWA 3 $\lambda/D$, OWA 9.1$\lambda/D$
\item Imaging in a wide field of view in Band 1 with an SPC---5.9 $\lambda/D$, OWA 20.1 $\lambda/D$
\end{enumerate}
The first mode is the instrument's only required mode, whereas the other 3 modes are `best effort'.  Conservative realizations of the modes are based directly on the results of thermal vacuum testing of the as-built instrument.  As this testing only required the instrument to perform to the level of its sole requirement, it did not establish the true limits of the instrument's performance.  The optimistic realizations are based on simulations of the instrument as designed, which predict static contrast capabilities beyond those established during the instrument's test campaign.  In addition to imaging and spectroscopy, \corgietc can also model polarimetric observations.

\begin{figure}[ht]
\centering
\includegraphics[width=\textwidth]{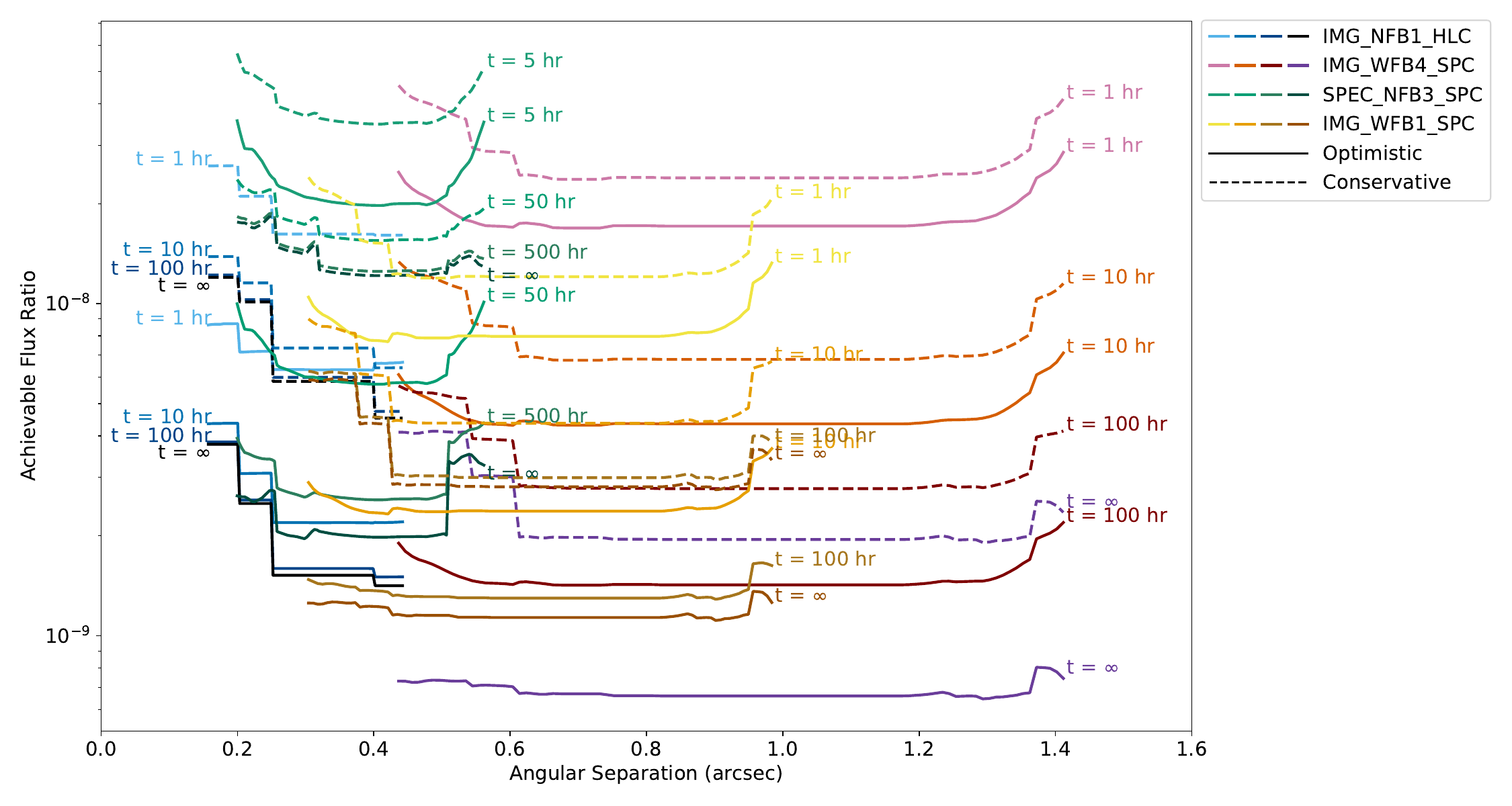}
\caption{Minimum achievable flux ratio at SNR=5 for a V=5 magnitude G5V star for various integration times for all observing modes currently supported by \corgietc.  Each set of colors represents varying integration times (in hours) for one observing mode (1, 10, 100, and $\infty$ for imaging modes and 5, 50, 500, $\infty$ for the spectroscopy mode).  Solid lines are optimistic scenarios and dashed lines are conservative ones. This figure was generated with the code available at \nlhref{https://github.com/roman-corgi/corgietc/blob/main/Notebooks/10_Contrast_Curves.ipynb} using \corgietc version 1.6.0 and EXOSIMS version 3.6.5.}\label{fig:contrast_curves}
\end{figure}

\reffig{fig:contrast_curves} shows the operation of \corgietc in answering questions 1 and 2 for all currently implemented observing modes. For each realization of each mode, we compute achievable flux ratio curves (sometimes known as contrast curves) for a 5 SNR detection of a point source about a 5th magnitude (V band) G5V star assuming a local zodiacal light contribution of 23 mag arcsec$^{-2}$ and an exozodiacal contribution equivalent to 1 solar system zodi for a planet located at a physical separation of 4.15 AU from its host star (for details on zodiacal light modeling, see \refnum{spohn2025refined}). For each imaging mode, curves were computed assuming 1, 10, and 100 hour integrations, with post-processing factors of 2 (that is, it was assumed that residual speckle could be suppressed by a factor of 2 via post-processing---see \refnum{nemati2023analytical} for details).  For the spectroscopy modes, curves were computed assuming 5, 50, and 500 hour integrations, with post-processing factors of 1.2. For each mode, a `saturation curve' was also computed, corresponding to the effective contrast achieved with infinite integration time.  This is effectively a measure of the instrument's contrast stability---the expected variance of the residual speckle field that cannot be modeled and would therefore not be removable via post-processing.

\begin{figure}[ht]
\centering
\includegraphics[width=0.495\textwidth]{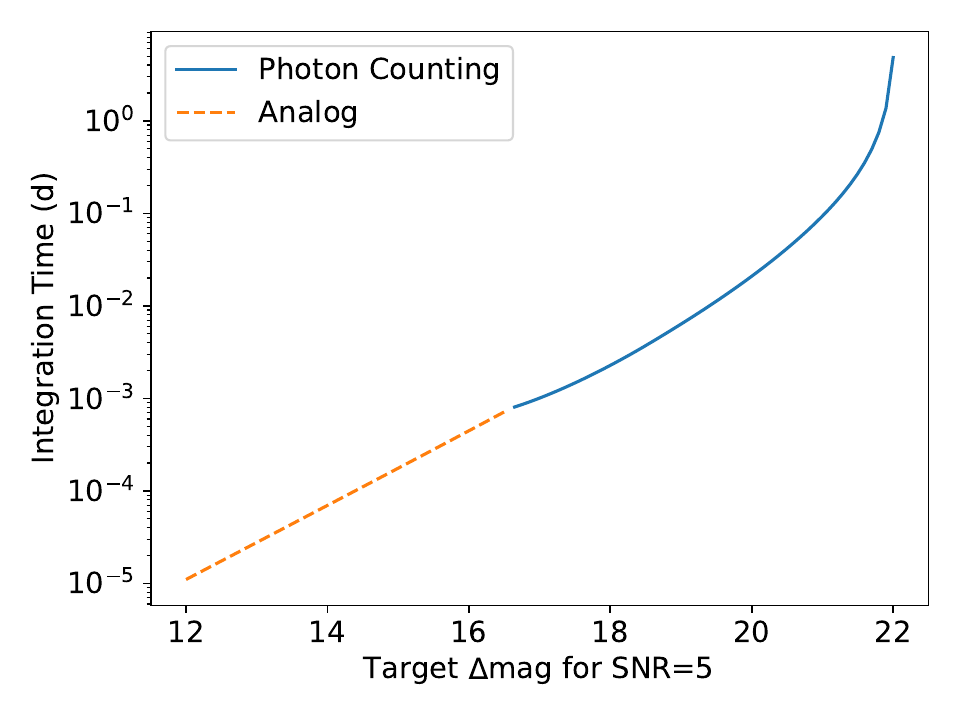}\hfill
\includegraphics[width=0.495\textwidth]{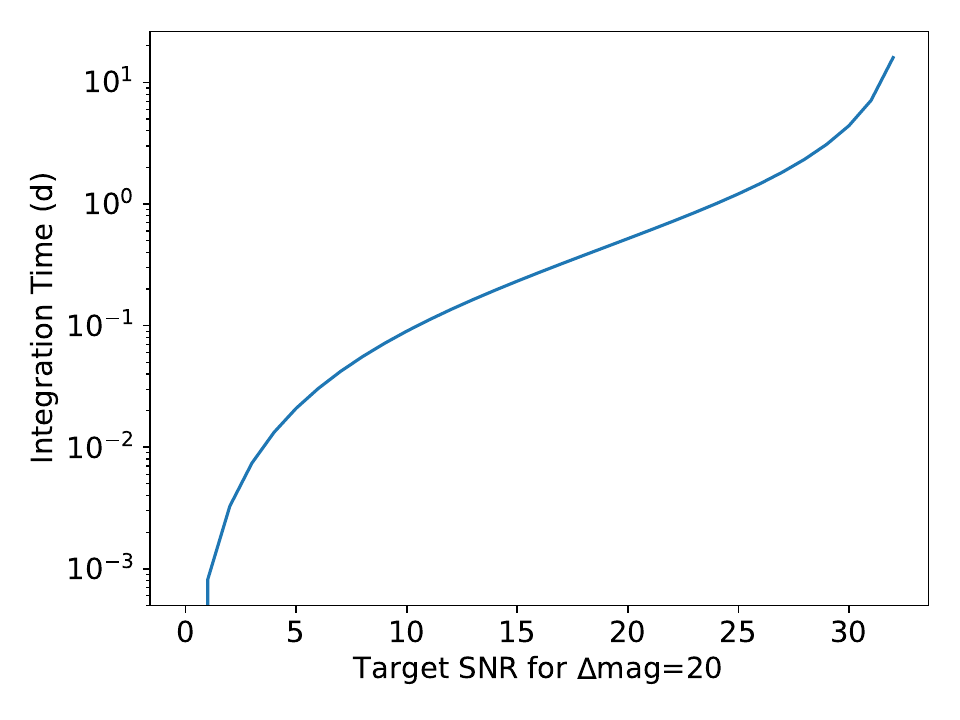}
\caption{For the same assumptions as in \reffig{fig:contrast_curves}: the integration times as a function of planet $\Delta$mag for a fixed SNR (\emph{left}) and as a function of required SNR for a fixed $\Delta$mag (\emph{right}).  In the left-hand figure, the operating mode of the detector (analog vs.~photon counting) is indicated via the line style.  This figure was generated with the code available at \nlhref{https://github.com/roman-corgi/corgietc/blob/main/Notebooks/01_Anatomy_of_an_Integration_Time_Calculation.ipynb} using \corgietc version 1.6.0 and EXOSIMS version 3.6.5.}\label{fig:intTime_curves}
\end{figure}

\reffig{fig:intTime_curves} shows the operation of \corgietc in answering question 3 for the required instrument mode (narrow-field, band 1, imaging with the HLC) with optimistic assumptions.  We assume the same synthetic target star and astrophysical noise sources as in \reffig{fig:contrast_curves} and plot the integration time required to reach an SNR of 5 under varying assumptions of planet $\Delta\textrm{mag} = -2.5\log_{10}F_r$.  We similarly plot the integration time required as a function of SNR for a fixed $\Delta\textrm{mag}$ of 20 (corresponding to a flux ratio of $10^{-8}$).  The coronagraph's Exoplanetary Systems Camera (EXCAM) can operate in two different modes: either as a photon counting device (in which case the read noise is effectively zero), or in a classic, `analog' mode, in which case the camera behaves more like a conventional, low read-noise, CCD.  Further details on the two modes are presented in \refnum{nemati2023analytical}.  The choice of which mode to use is driven primarily by the expected maximum and average fluxes on any pixel in any single frame, so as to avoid saturation of any pixel. By default, \corgietc automatically selects the operating mode based on estimates made at runtime of this flux, with a user-settable threshold value.  \reffig{fig:intTime_curves} shows the operation of this mode selection.  As the assumed planet flux grows, the expected flux value passes the threshold, and camera operation is set to analog mode.  If photon-counting mode were forced for all assumed values of $\Delta\textrm{mag}$, we would find the counter-intuitive result that the predicted integration time would actually start increasing for brighter planets due to detector effects.\footnote{Further details on this are provided in the \corgietc demonstration notebooks, packaged along with the code repository, and especially \nlhref{https://github.com/roman-corgi/corgietc/blob/main/Notebooks/01_Anatomy_of_an_Integration_Time_Calculation.ipynb}}. 

\section{Roman Coronagraph Pointing Tools}\label{sec:pointing}
\begin{figure}[ht]
\centering
\includegraphics[width=0.4\textwidth]{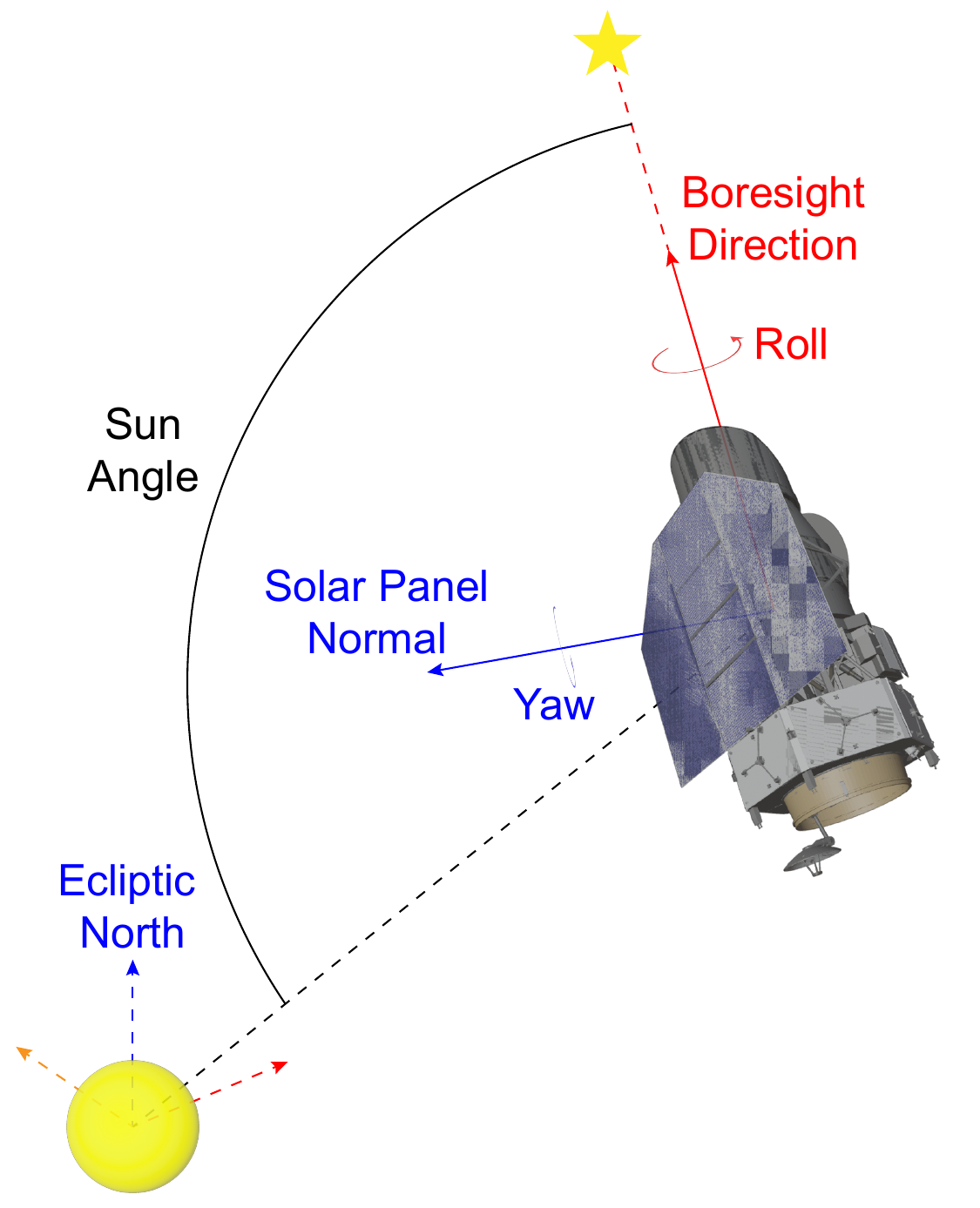}
\caption{Schematic of Roman body-fixed reference frame and pointing angles. The Roman 3D model was adapted from \nlhref{https://github.com/nasa/NASA-3D-Resources/}.}\label{fig:roman_pointing}
\end{figure}

In order to plan coronagraph observations, it is necessary to model the restrictions on observatory pointing, which include both restrictions on the overall observatory pointing as well as restrictions on relative pointing changes between target and reference stars.  \reffig{fig:roman_pointing} shows a schematic view of the observatory pointing model adopted in CPP pointing tools.\footnote{\nlhref{https://github.com/roman-corgi/roman_pointing}. Further details available at \nlhref{https://roman-pointing.readthedocs.io/en/latest}.}  The first unit direction of the Roman body-fixed reference frame is the boresight vector (the direction from telescope to target) and the third unit direction of the frame is the vector orthogonal to the solar panels.  The second unit direction is given by the cross product of the third and first directions (not shown for clarity, it is the direction into the page in the figure). Observatory pointing is given by a standard yaw-pitch-roll Euler angle set, with yaw representing counter-clockwise rotations about the third axis, pitch about the second axis, and roll about the first axis.  At zero ptich and roll, the solar panel normal direction points directly at the sun. 

The sun angle is defined as the angle between the boresight direction and the unit vector from the observatory to the sun.  This angle must be between 54$^\circ$ and 126$^\circ$ to ensure that the observatory is pointing sufficiently away from the sun to avoid glint and also such that the solar panels are sufficiently illuminated.  Similarly, solar panel illumination requirements restrict roll about the boresight to $\pm 15^\circ$. So as to avoid significant changes in observatory illumination, which can cause thermal transients impacting the optical system, the difference in pitch angle ($\Delta$pitch) between a target and reference star is limited to 5$^\circ$.  For more details on the modeling leading to this constraint, see \refnum{krist2023end}. It is important to note that this restriction is not a limit on the spherical angle on-sky between the pair of stars, as yaw remains unconstrained, meaning that it is possible to pair a target star in the ecliptic North hemisphere with one in the South, so long as the $\Delta$pitch constraint is met.  Operationally, it is best to avoid such pairings, as they lead to long slew times, but, given the small number of available reference stars, they are occasionally unavoidable. 

\begin{figure}[ht]
\centering
\includegraphics[width=\textwidth]{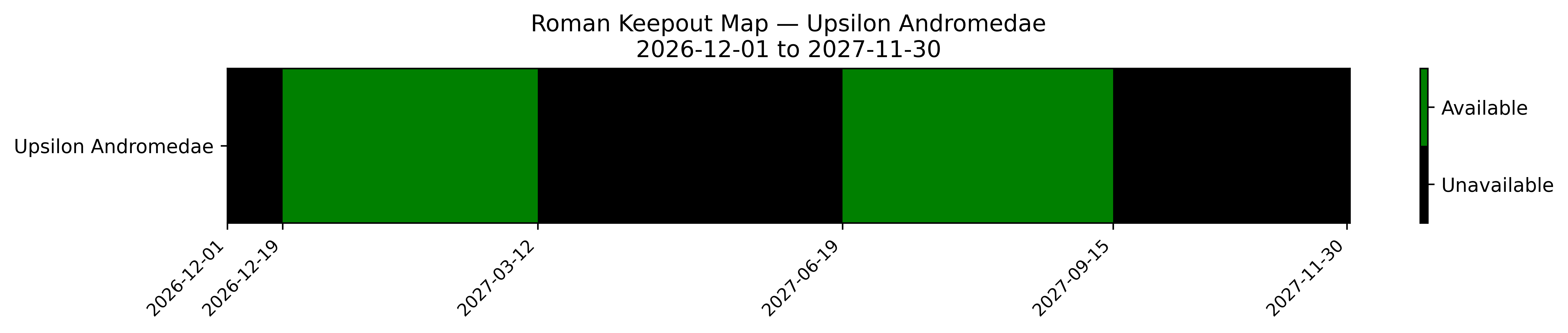}
\caption{Target availability for Upsilon Andromedae for the first year of nominal observing operations. This figure was generated with the code available at \nlhref{https://github.com/roman-corgi/roman_pointing/blob/main/Notebooks/}.}\label{fig:ups_and_keepout}
\end{figure}

\begin{figure}[ht]
\centering
\includegraphics[width=\textwidth, clip=true, trim=0pt 0pt 140pt 33pt]{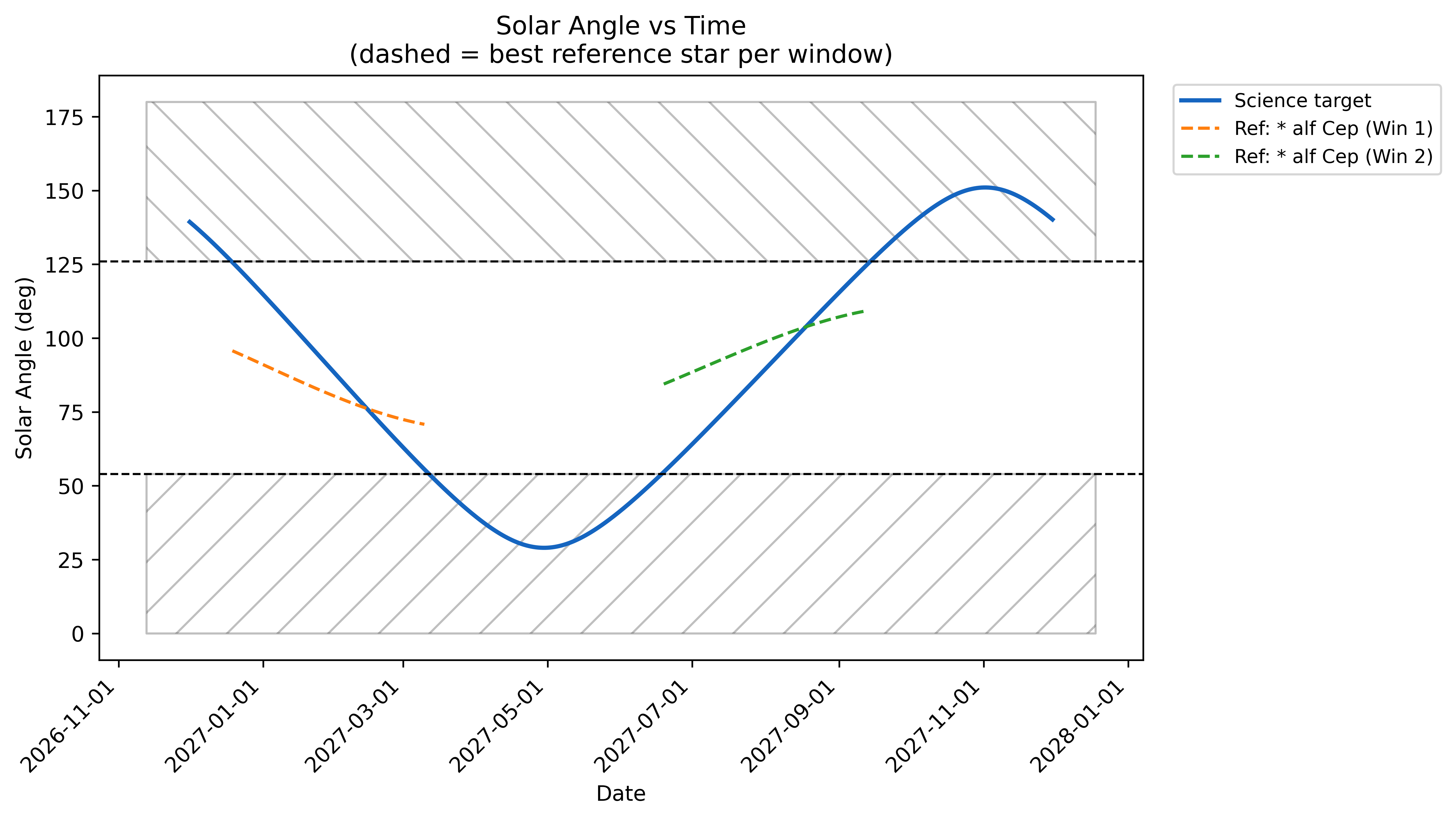}
\caption{Target and reference star availability for Upsilon Andromedae for the first year of nominal observing operations. The solid curve represents the sun angle of the target while the dashed curves represent the sun angles of the best available reference stars during the periods where the target is observable. This figure was generated with the code available at \nlhref{https://github.com/roman-corgi/roman_pointing/blob/main/Notebooks/}.}\label{fig:solar_angles}
\end{figure}

From the calculation of the sun angle and its associated restrictions, we can generate availability (or `keepout') maps for any target of interest.  \reffig{fig:ups_and_keepout} shows one such map for the high-priority reflected light target Upsilon Andromedae.  As this target lies below an absolute ecliptic  latitude of 54$^\circ$ (specifically at $\sim 30^\circ$ North), it is only observable for part of the year, in two roughly 3-month stretches. The target availability is overlaid on top of orbital modeling of known planets of interest in order to drive the optimal time for observations.   For each observable period, we can determine the `best' available reference star---the one chosen from the highest available tier (see \refnum{hom2026roman}) with the lowest magnitude in the observing band and with the lowest slew time from the target.  \reffig{fig:solar_angles} shows an example of reference star selection for Upsilon Andromedae.  In this case, Alpha Cephei is identified as the `best' reference star.

\section{CONCLUSIONS}
During its first 3 years of operations, the CPP has implemented an ecosystem of planning and modeling tools that are now being actively used to schedule the initial set of observations to be carried out with the Roman Coronagraph Instrument.  Assuming a launch close to the start of the launch readiness window, the first of these observations may take place as early as December 2026.  In addition to enabling planning activities for Roman, the developments described here have also led to a number of valuable lessons that should be considered for future direct imaging missions, and especially in the development of the Habitable Worlds Observatory (HWO).  The coronagraph's concept of operations reliance on reference stars represents a risk given the very small number of viable references, and the heavy scheduling constraints they impose on the observing program.  These constraints could potentially be even more stringent for HWO, where the target contrast levels might also necessitate matching references to targets by spectral type.  Detailed modeling of pitch angle impacts on internal coronagraph optics should be treated as a very high priority so that the least restrictive constraints may be placed on $\Delta$pitch.  Similarly, the overall observatory's field of regard and sun angle restrictions play a very important role in determining mission capabilities, and should be included as part of the early trade space in designing the observatory.  Studies have already been undertaken to evaluate these effects for HWO (see, e.g., \refnum{spohn2026understanding}) and have clearly demonstrated that overall mission yield can be highly dependent on field of regard. The most valuable lessons to come from the Roman Coronagraph, however, will be those learned from on-sky operations.  The data collected on orbit will demonstrate, for the first time, how to properly structure a space-based direct imaging campaign, and will directly inform the mission design for HWO and all future planet imagers. 

\acknowledgments 
This work was supported by NASA under award No. 80NSSC24K0216. The research was carried out in part through the Jet Propulsion Laboratory, California Institute of Technology, under a contract with the National Aeronautics and Space Administration (80NM0018D0004). The CPP gratefully acknowledges the contributions of the $>1600$ scientists and engineers that have helped make the Coronagraph Instrument possible. This research has made use of the SIMBAD database, operated at CDS, Strasbourg, France.\cite{wenger2000simbad} This research has made use of the NASA Exoplanet Archive, which is operated by the California Institute of Technology, under contract with the National Aeronautics and Space Administration under the Exoplanet Exploration Program.\cite{christiansen2025NASA}   This work made use of Astropy (https://www.astropy.org) a community-developed core Python package and an ecosystem of tools and resources for astronomy.\cite{astropy:2013, astropy:2018, astropy:2022}

\bibliography{Main}   

\begin{thebibliography}{10}

\bibitem{spergel2015wide}
{Spergel}, D., {Gehrels}, N., {Baltay}, C., {Bennett}, D., {Breckinridge}, J.,
  {Donahue}, M., {Dressler}, A., {Gaudi}, B.~S., {Greene}, T., {Guyon}, O.,
  {Hirata}, C., {Kalirai}, J., {Kasdin}, N.~J., {Macintosh}, B., {Moos}, W.,
  {Perlmutter}, S., {Postman}, M., {Rauscher}, B., {Rhodes}, J., {Wang}, Y.,
  {Weinberg}, D., {Benford}, D., {Hudson}, M., {Jeong}, W.-S., {Mellier}, Y.,
  {Traub}, W., {Yamada}, T., {Capak}, P., {Colbert}, J., {Masters}, D.,
  {Penny}, M., {Savransky}, D., {Stern}, D., {Zimmerman}, N., {Barry}, R.,
  {Bartusek}, L., {Carpenter}, K., {Cheng}, E., {Content}, D., {Dekens}, F.,
  {Demers}, R., {Grady}, K., {Jackson}, C., {Kuan}, G., {Kruk}, J., {Melton},
  M., {Nemati}, B., {Parvin}, B., {Poberezhskiy}, I., {Peddie}, C., {Ruffa},
  J., {Wallace}, J.~K., {Whipple}, A., {Wollack}, E., and {Zhao}, F.,
  ``{Wide-Field InfrarRed Survey Telescope-Astrophysics Focused Telescope
  Assets WFIRST-AFTA 2015 Report},'' {\em arXiv preprint arXiv:1503.03757}
  (2015).

\bibitem{akeson2019wide}
Akeson, R., Armus, L., Bachelet, E., Bailey, V., Bartusek, L., Bellini, A.,
  Benford, D., Bennett, D., Bhattacharya, A., Bohlin, R., et~al., ``The wide
  field infrared survey telescope: 100 hubbles for the 2020s,'' {\em arXiv
  preprint arXiv:1902.05569}  (2019).

\bibitem{douglas2020review}
{Douglas}, E.~S., {Ashcraft}, J.~N., {Belikov}, R., {Debes}, J., {Kasdin}, J.,
  {Krist}, J., {Lacy}, B.~I., {Nemati}, B., {Milani}, K., {Pogorelyuk}, L.,
  {Riggs}, A.~J.~E., {Savransky}, D., and {Sirbu}, D., ``{A review of
  simulation and performance modeling tools for the Roman coronagraph
  instrument},'' in [{\em Space Telescopes and Instrumentation 2020: Optical,
  Infrared, and Millimeter Wave}{\nolinebreak\hspace{0.1em}]},  Lystrup, M.,
  Perrin, M.~D., Batalha, N., Siegler, N., and Tong, E.~C., eds., {\em Society
  of Photo-Optical Instrumentation Engineers (SPIE) Conference Series} {\bf
  11443},  1144338, Society of Photo-Optical Instrumentation Engineers, SPIE
  (Dec. 2020).

\bibitem{kuan2025roman}
Kuan, G.~M., Monacelli, B., Baker, C., Tang, H., Rodgers, M., Willems, P.,
  Marx, D., Rupp, J., Groff, T., Kern, B., Nemati, B., Colavita, M., and
  Poberezhskiy, I., ``{Roman coronagraph instrument optical design
  description},'' {\em Journal of Astronomical Telescopes, Instruments, and
  Systems}~{\bf 11}(2),  021413 (2025).

\bibitem{krist2023end}
Krist, J.~E., Steeves, J.~B., Dube, B.~D., Eldorado~Riggs, A., Kern, B.~D.,
  Marx, D.~S., Cady, E.~J., Zhou, H., Poberezhskiy, I.~Y., Baker, C.~W.,
  et~al., ``End-to-end numerical modeling of the roman space telescope
  coronagraph,'' {\em Journal of Astronomical Telescopes, Instruments, and
  Systems}~{\bf 9}(4),  045002--045002 (2023).

\bibitem{riggs2025flight}
Riggs, A.~E., Bailey, V.~P., Moody, D., Balasubramanian, K., Basinger, S.~A.,
  Belikov, R., Bendek, E., Debes, J., Dube, B.~D., Gersh-Range, J., et~al.,
  ``Flight masks of the roman space telescope coronagraph instrument,'' {\em
  Journal of Astronomical Telescopes, Instruments, and Systems}~{\bf 11}(2),
  021403--021403 (2025).

\bibitem{groff2025spectroscopy}
Groff, T.~D., Zimmerman, N.~T., Bray, E., Woodland, M.~N., Subedi, H.~B., III,
  J. J.~L., Gao, G., Titus, J., Bell, D., Bailey, V.~P., Baker, C.,
  Hildebrandt, S., Monacelli, B., Yamada, T., and Tamura, M., ``{Spectroscopy
  and polarimetry design and flight instrument calibration for the Roman
  Coronagraph},'' {\em Journal of Astronomical Telescopes, Instruments, and
  Systems}~{\bf 11}(3),  031510 (2025).

\bibitem{nemati2023analytical}
Nemati, B., Krist, J., Poberezhskiy, I., and Kern, B., ``Analytical performance
  model and error budget for the roman coronagraph instrument,'' {\em Journal
  of Astronomical Telescopes, Instruments, and Systems}~{\bf 9}(3),
  034007--034007 (2023).

\bibitem{poberezhskiy2025overview}
Poberezhskiy, I.~Y., Cady, E.~J., Heydorff, K., Kern, B., Luchik, T.~S., Zhao,
  F., Bailey, V.~P., Bush, N., Colavita, M.~M., Creager, B., Fathpour, N.,
  Gaidon, C., Grue, A., Kempenaar, J.~E., Krist, J.~E., Kuan, G.~M., Lam,
  J.~C., Mandi{\'c}, M., Marx, D.~S., Nemati, B., Riggs, A. J.~E., Seo, B.-J.,
  Shi, F., Smith, M.~W., and Zhou, H., ``{Overview of Roman Coronagraph
  Instrument requirements, test campaign, and results},'' {\em Journal of
  Astronomical Telescopes, Instruments, and Systems}~{\bf 11}(3),  031511
  (2025).

\bibitem{cady2025high}
Cady, E., Bowman, N., Greenbaum, A.~Z., Ingalls, J.~G., Kern, B., Krist, J.,
  Marx, D., Poberezhskiy, I., Riggs, A. J.~E., Ruane, G., Seo, B.-J., Shi, F.,
  and Zhou, H., ``{High-order wavefront sensing and control for the Roman
  Coronagraph Instrument (CGI): architecture and measured performance},'' {\em
  Journal of Astronomical Telescopes, Instruments, and Systems}~{\bf 11}(2),
  021408 (2025).

\bibitem{zhou2025high}
Zhou, H., Cady, E., Kern, B., Krist, J., Marx, D., Poberezhskiy, I., Riggs,
  A.~E., Ruane, G., Seo, B.-J., and Shi, F., ``{High-order wavefront sensing
  and control performance modeling and model validations with Roman coronagraph
  instrument},'' {\em Journal of Astronomical Telescopes, Instruments, and
  Systems}~{\bf 11}(2),  021410 (2025).

\bibitem{wolff2024roman}
Wolff, S.~G., Wang, J., Stapelfeldt, K., Bailey, V.~P., Savransky, D., Hom, J.,
  Biller, B., Brandner, W., Anche, R., Blunt, S., Brinjikji, M., Girard, J.~H.,
  Krause, O., Li, Z., Livingston, J., Millar-Blanchaer, M.~A., Noel, M., Pueyo,
  L., Rosa, R. J.~D., Samland, M., and Schragal, N., ``{The Roman coronagraph
  community participation program: observation planning},'' in [{\em Space
  Telescopes and Instrumentation 2024: Optical, Infrared, and Millimeter
  Wave}{\nolinebreak\hspace{0.1em}]},  Coyle, L.~E., Matsuura, S., and Perrin,
  M.~D., eds.,  {\bf 13092},  1309255, International Society for Optics and
  Photonics, SPIE (2024).

\bibitem{wolff2026roman}
Wolff, S.~G., Bailey, V.~P., Hom, J., Millar-Blanchaer, M.~A., Redmond, S.~F.,
  Girard, J.~H., Carri{\'o}n-Gonz{\'a}lez, O., Savransky, D., Mazoyer, J.,
  Vega-Pallauta, M.~C., Chauvin, G., Ravet, M., Greenbaum, A.~Z., Cady, E.~J.,
  and Anche, R.~M., ``{The Roman Coronagraph Community Participation Program:
  trials and triumphs of designing an observing program for a technology
  demonstration instrument},'' in [{\em Space Telescopes and Instrumentation
  2026: Optical, Infrared, and Millimeter Wave}{\nolinebreak\hspace{0.1em}]},
  Coyle, L.~E., Matsuura, S., and Perrin, M.~D., eds., International Society
  for Optics and Photonics, SPIE (2026).

\bibitem{hom2026roman}
Hom, J. et~al., ``{The Roman Coronagraph Community Participation Program:
  pre-launch reference star list and impact of reference star properties on
  post-processing performance},'' in [{\em Space Telescopes and Instrumentation
  2026: Optical, Infrared, and Millimeter Wave}{\nolinebreak\hspace{0.1em}]},
  Coyle, L.~E., Matsuura, S., and Perrin, M.~D., eds., International Society
  for Optics and Photonics, SPIE (2026).

\bibitem{savransky2024roman}
Savransky, D., Bailey, V.~P., Wolff, S.~G., Millar-Blanchaer, M.~A., Wang, J.,
  Altinier, L., Anche, R., Baudoz, P., Biller, B., Blunt, S., Brandner, W.,
  Brinjikji, M., Carri{\'o}n-Gonz{\'a}lez, O., Chavez, A., Choquet, E.,
  Doelman, D., Girard, J.~H., Greenbaum, A.~Z., Hasler, S.~N., Hom, J.,
  Ingalls, J.~G., Kane, S.~R., Kasdin, N.~J., Krause, O., Kuzuhara, M., Lau,
  A., Li, Z., Livingston, J., Lowrance, P.~J., Ludwick, K., Macintosh, B.,
  Mamajek, E., Marley, M., Mazoyer, J., Mennesson, B., Mizuki, T., Moran,
  S.~E., Murakami, N., Nishikawa, J., Noel, M., Pueyo, L., Rafels, S.~H.,
  Rhodes, J., Robinson, T., Rosa, R. J.~D., Samland, M., Schragal, N.,
  Schreiber, J., Sobeck, J., Stapelfeldt, K., Tamura, M., Uyajma, T., Vigan,
  A., Woodland, M., Ygouf, M., Yoneta, K., Zellem, R.~T., and Zimmerman, N.~T.,
  ``{The Nancy Grace Roman Space Telescope coronagraph community participation
  program},'' in [{\em Space Telescopes and Instrumentation 2024: Optical,
  Infrared, and Millimeter Wave}{\nolinebreak\hspace{0.1em}]},  Coyle, L.~E.,
  Matsuura, S., and Perrin, M.~D., eds.,  {\bf 13092},  130921I, International
  Society for Optics and Photonics, SPIE (2024).

\bibitem{wang2026roman}
Wang, J.~J. et~al., ``{The Roman Coronagraph Community Participation Program:
  data reduction pipeline design and implementation},'' in [{\em Space
  Telescopes and Instrumentation 2026: Optical, Infrared, and Millimeter
  Wave}{\nolinebreak\hspace{0.1em}]},  Coyle, L.~E., Matsuura, S., and Perrin,
  M.~D., eds., International Society for Optics and Photonics, SPIE (2026).

\bibitem{savransky2019exploration}
{Savransky}, D., {Gasc{\'o}n}, C., {Kinzly}, N., {Batalha}, N., {Lewis}, N.,
  and {Marley}, M., ``{Exploration of the dynamical phase space of stars with
  known planets},'' in [{\em Techniques and Instrumentation for Detection of
  Exoplanets IX}{\nolinebreak\hspace{0.1em}]},  Shaklan, S.~B., ed., {\em
  Society of Photo-Optical Instrumentation Engineers (SPIE) Conference Series}
  {\bf 11117},  111171D (Sept. 2019).

\bibitem{brown2005}
Brown, R.~A., ``Single-visit photometric and obscurational completeness,'' {\em
  The Astrophysical Journal}~{\bf 624},  1010--1024 (2005).

\bibitem{savransky2017exosims}
{Savransky}, D., {Delacroix}, C., and {Garrett}, D., ``{EXOSIMS: Exoplanet
  Open-Source Imaging Mission Simulator}.'' Astrophysics Source Code Library (6
  2017).

\bibitem{spohn2025refined}
Spohn, C., Savransky, D., and Stark, C.~C., ``A refined photometric constraint
  for exoplanet direct imaging yield estimation and observation scheduling,''
  {\em The Astronomical Journal}~{\bf 170}(6),  354 (2025).

\bibitem{spohn2026understanding}
Spohn, C., Stark, C.~C., Savransky, D., and Latouf, N., ``{Understanding HWO's
  field of regard and characterization requirement trade space with a dynamic
  observation scheduling algorithm},'' {\em Journal of Astronomical Telescopes,
  Instruments, and Systems}~{\bf 12}(4),  041010 (2026).

\bibitem{wenger2000simbad}
Wenger, M., Ochsenbein, F., Egret, D., Dubois, P., Bonnarel, F., Borde, S.,
  Genova, F., Jasniewicz, G., Lalo{\"e}, S., Lesteven, S., et~al., ``The simbad
  astronomical database-the cds reference database for astronomical objects,''
  {\em Astronomy and Astrophysics Supplement Series}~{\bf 143}(1),  9--22
  (2000).

\bibitem{christiansen2025NASA}
Christiansen, J.~L., McElroy, D.~L., Harbut, M., Ciardi, D.~R., Crane, M.,
  Good, J., Hardegree-Ullman, K.~K., Kesseli, A.~Y., Lund, M.~B., Lynn, M.,
  Muthiar, A., Nilsson, R., Oluyide, T., Papin, M., Rivera, A., Swain, M.,
  Susemiehl, N.~D., Tam, R., van Eyken, J., and Beichman, C., ``The nasa
  exoplanet archive and exoplanet follow-up observing program: Data, tools, and
  usage,'' {\em The Planetary Science Journal}~{\bf 6},  186 (aug 2025).

\bibitem{astropy:2013}
{Astropy Collaboration}, {Robitaille}, T.~P., {Tollerud}, E.~J., {Greenfield},
  P., {Droettboom}, M., {Bray}, E., {Aldcroft}, T., {Davis}, M., {Ginsburg},
  A., {Price-Whelan}, A.~M., {Kerzendorf}, W.~E., {Conley}, A., {Crighton}, N.,
  {Barbary}, K., {Muna}, D., {Ferguson}, H., {Grollier}, F., {Parikh}, M.~M.,
  {Nair}, P.~H., {Unther}, H.~M., {Deil}, C., {Woillez}, J., {Conseil}, S.,
  {Kramer}, R., {Turner}, J.~E.~H., {Singer}, L., {Fox}, R., {Weaver}, B.~A.,
  {Zabalza}, V., {Edwards}, Z.~I., {Azalee Bostroem}, K., {Burke}, D.~J.,
  {Casey}, A.~R., {Crawford}, S.~M., {Dencheva}, N., {Ely}, J., {Jenness}, T.,
  {Labrie}, K., {Lim}, P.~L., {Pierfederici}, F., {Pontzen}, A., {Ptak}, A.,
  {Refsdal}, B., {Servillat}, M., and {Streicher}, O., ``{Astropy: A community
  Python package for astronomy},'' {\em Astronomy \& Astrophysics}~{\bf 558},
  A33 (10 2013).

\bibitem{astropy:2018}
{Price-Whelan}, A.~M., {Sip{\H{o}}cz}, B.~M., {G{\"u}nther}, H.~M., {Lim},
  P.~L., {Crawford}, S.~M., {Conseil}, S., {Shupe}, D.~L., {Craig}, M.~W.,
  {Dencheva}, N., {Ginsburg}, A., {VanderPlas}, J.~T., {Bradley}, L.~D.,
  {P{\'e}rez-Su{\'a}rez}, D., {de Val-Borro}, M., {Paper Contributors}, P.,
  {Aldcroft}, T.~L., {Cruz}, K.~L., {Robitaille}, T.~P., {Tollerud}, E.~J.,
  {Coordination Committee}, A., {Ardelean}, C., {Babej}, T., {Bach}, Y.~P.,
  {Bachetti}, M., {Bakanov}, A.~V., {Bamford}, S.~P., {Barentsen}, G.,
  {Barmby}, P., {Baumbach}, A., {Berry}, K.~L., {Biscani}, F., {Boquien}, M.,
  {Bostroem}, K.~A., {Bouma}, L.~G., {Brammer}, G.~B., {Bray}, E.~M.,
  {Breytenbach}, H., {Buddelmeijer}, H., {Burke}, D.~J., {Calderone}, G., {Cano
  Rodr{\'\i}guez}, J.~L., {Cara}, M., {Cardoso}, J.~V.~M., {Cheedella}, S.,
  {Copin}, Y., {Corrales}, L., {Crichton}, D., {D{\textquoteright}Avella}, D.,
  {Deil}, C., {Depagne}, {\'E}., {Dietrich}, J.~P., {Donath}, A., {Droettboom},
  M., {Earl}, N., {Erben}, T., {Fabbro}, S., {Ferreira}, L.~A., {Finethy}, T.,
  {Fox}, R.~T., {Garrison}, L.~H., {Gibbons}, S.~L.~J., {Goldstein}, D.~A.,
  {Gommers}, R., {Greco}, J.~P., {Greenfield}, P., {Groener}, A.~M.,
  {Grollier}, F., {Hagen}, A., {Hirst}, P., {Homeier}, D., {Horton}, A.~J.,
  {Hosseinzadeh}, G., {Hu}, L., {Hunkeler}, J.~S., {Ivezi{\'c}}, {\v{Z}}.,
  {Jain}, A., {Jenness}, T., {Kanarek}, G., {Kendrew}, S., {Kern}, N.~S.,
  {Kerzendorf}, W.~E., {Khvalko}, A., {King}, J., {Kirkby}, D., {Kulkarni},
  A.~M., {Kumar}, A., {Lee}, A., {Lenz}, D., {Littlefair}, S.~P., {Ma}, Z.,
  {Macleod}, D.~M., {Mastropietro}, M., {McCully}, C., {Montagnac}, S.,
  {Morris}, B.~M., {Mueller}, M., {Mumford}, S.~J., {Muna}, D., {Murphy},
  N.~A., {Nelson}, S., {Nguyen}, G.~H., {Ninan}, J.~P., {N{\"o}the}, M.,
  {Ogaz}, S., {Oh}, S., {Parejko}, J.~K., {Parley}, N., {Pascual}, S., {Patil},
  R., {Patil}, A.~A., {Plunkett}, A.~L., {Prochaska}, J.~X., {Rastogi}, T.,
  {Reddy Janga}, V., {Sabater}, J., {Sakurikar}, P., {Seifert}, M., {Sherbert},
  L.~E., {Sherwood-Taylor}, H., {Shih}, A.~Y., {Sick}, J., {Silbiger}, M.~T.,
  {Singanamalla}, S., {Singer}, L.~P., {Sladen}, P.~H., {Sooley}, K.~A.,
  {Sornarajah}, S., {Streicher}, O., {Teuben}, P., {Thomas}, S.~W., {Tremblay},
  G.~R., {Turner}, J.~E.~H., {Terr{\'o}n}, V., {van Kerkwijk}, M.~H., {de la
  Vega}, A., {Watkins}, L.~L., {Weaver}, B.~A., {Whitmore}, J.~B., {Woillez},
  J., {Zabalza}, V., and {Contributors}, A., ``{The Astropy Project: Building
  an Open-science Project and Status of the v2.0 Core Package},'' {\em The
  Astronomical Journal}~{\bf 156},  123 (9 2018).

\bibitem{astropy:2022}
{Astropy Collaboration}, {Price-Whelan}, A.~M., {Lim}, P.~L., {Earl}, N.,
  {Starkman}, N., {Bradley}, L., {Shupe}, D.~L., {Patil}, A.~A., {Corrales},
  L., {Brasseur}, C.~E., {N{\"o}the}, M., {Donath}, A., {Tollerud}, E.,
  {Morris}, B.~M., {Ginsburg}, A., {Vaher}, E., {Weaver}, B.~A., {Tocknell},
  J., {Jamieson}, W., {van Kerkwijk}, M.~H., {Robitaille}, T.~P., {Merry}, B.,
  {Bachetti}, M., {G{\"u}nther}, H.~M., {Aldcroft}, T.~L., {Alvarado-Montes},
  J.~A., {Archibald}, A.~M., {B{\'o}di}, A., {Bapat}, S., {Barentsen}, G.,
  {Baz{\'a}n}, J., {Biswas}, M., {Boquien}, M., {Burke}, D.~J., {Cara}, D.,
  {Cara}, M., {Conroy}, K.~E., {Conseil}, S., {Craig}, M.~W., {Cross}, R.~M.,
  {Cruz}, K.~L., {D'Eugenio}, F., {Dencheva}, N., {Devillepoix}, H. A.~R.,
  {Dietrich}, J.~P., {Eigenbrot}, A.~D., {Erben}, T., {Ferreira}, L.,
  {Foreman-Mackey}, D., {Fox}, R., {Freij}, N., {Garg}, S., {Geda}, R.,
  {Glattly}, L., {Gondhalekar}, Y., {Gordon}, K.~D., {Grant}, D., {Greenfield},
  P., {Groener}, A.~M., {Guest}, S., {Gurovich}, S., {Handberg}, R., {Hart},
  A., {Hatfield-Dodds}, Z., {Homeier}, D., {Hosseinzadeh}, G., {Jenness}, T.,
  {Jones}, C.~K., {Joseph}, P., {Kalmbach}, J.~B., {Karamehmetoglu}, E.,
  {Ka{\l}uszy{\'n}ski}, M., {Kelley}, M. S.~P., {Kern}, N., {Kerzendorf},
  W.~E., {Koch}, E.~W., {Kulumani}, S., {Lee}, A., {Ly}, C., {Ma}, Z.,
  {MacBride}, C., {Maljaars}, J.~M., {Muna}, D., {Murphy}, N.~A., {Norman}, H.,
  {O'Steen}, R., {Oman}, K.~A., {Pacifici}, C., {Pascual}, S.,
  {Pascual-Granado}, J., {Patil}, R.~R., {Perren}, G.~I., {Pickering}, T.~E.,
  {Rastogi}, T., {Roulston}, B.~R., {Ryan}, D.~F., {Rykoff}, E.~S., {Sabater},
  J., {Sakurikar}, P., {Salgado}, J., {Sanghi}, A., {Saunders}, N.,
  {Savchenko}, V., {Schwardt}, L., {Seifert-Eckert}, M., {Shih}, A.~Y., {Jain},
  A.~S., {Shukla}, G., {Sick}, J., {Simpson}, C., {Singanamalla}, S., {Singer},
  L.~P., {Singhal}, J., {Sinha}, M., {Sip{\H{o}}cz}, B.~M., {Spitler}, L.~R.,
  {Stansby}, D., {Streicher}, O., {{\v{S}}umak}, J., {Swinbank}, J.~D.,
  {Taranu}, D.~S., {Tewary}, N., {Tremblay}, G.~R., {de Val-Borro}, M., {Van
  Kooten}, S.~J., {Vasovi{\'c}}, Z., {Verma}, S., {de Miranda Cardoso}, J.~V.,
  {Williams}, P. K.~G., {Wilson}, T.~J., {Winkel}, B., {Wood-Vasey}, W.~M.,
  {Xue}, R., {Yoachim}, P., {Zhang}, C., {Zonca}, A., and {Astropy Project
  Contributors}, ``{The Astropy Project: Sustaining and Growing a
  Community-oriented Open-source Project and the Latest Major Release (v5.0) of
  the Core Package},'' {\em The Astrophysical Journal}~{\bf 935},  167 (Aug.
  2022).

\end{thebibliography}
\bibliographystyle{spiebib}

\end{document}